\UseRawInputEncoding
\documentclass[aip,graphicx]{revtex4-1}
\usepackage{graphicx}

\draft 

\begin{document}


\title{Experimental Upper Bounds for Resonance-Enhanced Entangled Two-Photon Absorption Cross Section of Indocyanine Green} 



\author{Manni He}
\author{Bryce P. Hickam}
\author{Nathan Harper}
\author{Scott K. Cushing}
 \email{scushing@caltech.edu.}
\affiliation{ 
Division of Chemistry and Chemical Engineering,
California Institute of Technology, Pasadena, CA 91125, USA
}%


\date{\today}

\begin{abstract}
Resonant intermediate states have been proposed to increase the efficiency of entangled two-photon absorption (ETPA). Although resonance-enhanced ETPA (r-ETPA) has been demonstrated in atomic systems using bright squeezed vacuum, it has not been studied in organic molecules. We investigate for the first time r-ETPA in an organic molecular dye, indocyanine green (ICG), when excited by broadband entangled photons in near-IR. Similar to many reported virtual state mediated ETPA (v-ETPA) measurements, no r-ETPA signals are measured, with an experimental upper bound for the cross section placed at $6 \times 10^{-23}$ cm$^2$/molecule. In addition, the classical resonance-enhanced two-photon absorption (r-TPA) cross section of ICG at 800 nm is measured for the first time to be \(20(\pm13)\) GM, suggesting that having a resonant intermediate state does not significantly enhance two-photon processes in ICG. The spectrotemporally resolved emission signatures of ICG excited by entangled photons are also presented to support this conclusion.
\end{abstract}


\maketitle 

\section{Introduction}

Research on molecular entangled two-photon absorption (ETPA) has gained momentum over the past decade due to its potential application in nonlinear spectroscopy and bioimaging. While multiple theoretical predictions on ETPA cross sections have been made \cite{landes_quantifying_2021, schlawin_polarization-entangled_2022, kang_efficient_2020, burdick_predicting_2018, wittkop_multichromatic_2023}, the predicted values vary by orders of magnitude, and the formulations emphasize different parameters of the excitation flux. Although reported ETPA cross sections could be as high as $1 \times 10^{-17}$ cm$^2$/molecule \cite{harpham_thiophene_2009, guzman_spatial_2010, upton_optically_2013}, recent measurements of ETPA cross sections reported values lower than $1 \times 10^{-21}$ cm$^2$/molecule for organic molecules with virtual intermediate states, even in dyes with near-unity quantum yields which facilitate fluorescence detection such as rhodamine 6G (R6G) \cite{tabakaev_spatial_2022, parzuchowski_setting_2021, landes_experimental_2021-1, mikhaylov_hot-band_2022, hickam_single-photon_2022}, and quantum dot systems with large classical two-photon absorption (TPA) cross sections \cite{gabler_photon_2023}. The measured fluorescence signals from ETPA are usually in the tens per second to tens per hour count rate range, which hinders many potential applications in imaging and sensing. 

For classical TPA, it is known that a real intermediate state increases the cross section by 1-2 orders of magnitude \cite{terenziani_enhanced_2008, kobayashi_stepwise_2018, drobizhev_resonance_2002, makarov_resonance_2007, hahn_two-step_2021}. Resonance-enhanced ETPA (r-ETPA) is predicted to be enhanced by similar orders of magnitude in atomic systems \cite{oka_two-photon_2018, kang_efficient_2020, schlawin_theory_2017, li_enhancement_2023}. While r-ETPA in atomic cesium \cite{georgiades_nonclassical_1995} and atomic rubidium \cite{dayan_two_2004} have been observed when excited by bright squeezed vacuum, the quantum enhancement on the resonance-enhanced TPA (r-TPA) cross sections was not quantified. Moreover, atoms generally have classical TPA cross sections in the range of $10^{-40} - 10^{-35}$ cm$^4$ s, which are 7-15 orders of magnitude larger than those of organic molecules ($10^{-50} - 10^{-47}$ cm$^4$ s) \cite{sulham_blue_2010, davila_two-photon_2018, bjorkholm_resonant_1974, xu_recent_2020, terenziani_enhanced_2008, nociarova_direct_2021}. Atoms also have longer-lived excited state coherences than molecules \cite{deutsch_spin_2010, boyd_optical_2006, young_half-minute-scale_2020, gustin_mapping_2023, hildner_femtosecond_2011, paulus_leveraging_2020} which could facilitate coherent biphoton processes. To date, experimental studies in organic molecules have only focused on virtual state-mediated ETPA (v-ETPA) instead of r-ETPA processes \cite{lee_entangled_2006, harpham_thiophene_2009, guzman_spatial_2010, upton_optically_2013, villabona-monsalve_entangled_2017, villabona-monsalve_two-photon_2018, eshun_investigations_2018, villabona-monsalve_measurements_2020, tabakaev_energy-time-entangled_2021, varnavski_two-photon_2020, parzuchowski_setting_2021, landes_quantifying_2021, landes_experimental_2021, mikhaylov_hot-band_2022, burdick_enhancing_2021, corona-aquino_experimental_2022, tabakaev_spatial_2022, villabona-monsalve_two-photon_2022, triana-arango_spectral_2023, varnavski_colors_2023}.  

\begin{figure}[h]
\centering
\includegraphics[scale = 0.55]{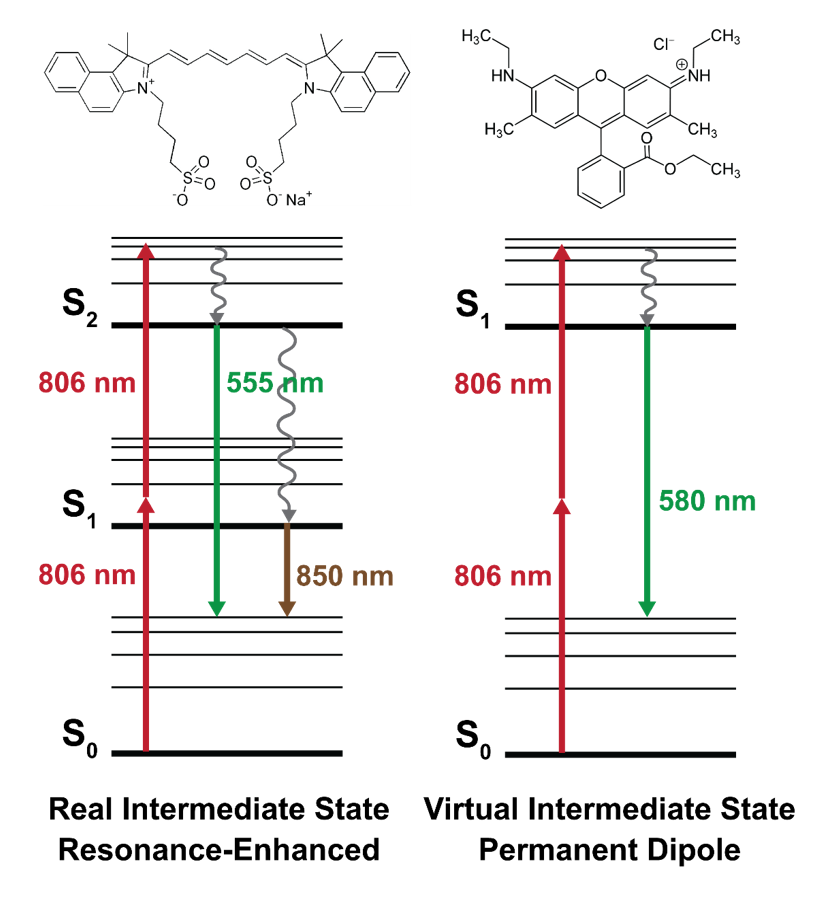}%
\caption{Jablonski diagram comparison between r-ETPA in ICG (left) and v-ETPA in R6G (right). Emission wavelengths are measured from 1 mM solutions in dimethyl sulfoxide (DMSO).
\label{fig:jablonski-diagrams}
}
\end{figure}

The experimental challenge in measuring molecular r-ETPA is that signals from the first singlet excited state (S1) can mask signals from the subsequent, weaker excitation to the second singlet excited state (S2). Traditional TPA measurement approaches such as transmission measurements, pump versus pairs power attenuation test \cite{tabakaev_spatial_2022, mikhaylov_hot-band_2022, dayan_nonlinear_2005}, and z-scans \cite{sengupta_sensitive_2000, ajami_two-photon_2010, nag_two-photon_2009} therefore become more difficult. Indocyanine green (ICG), however, is a unique case for r-ETPA because it has both an S1-S0 emission at $\sim$850 nm and an S2-S0 emission at $\sim$550 nm \cite{kumari_two-photon_2019, das_polarization_2013, pu_enhancing_2014}, as shown in \textbf{Figure \ref{fig:jablonski-diagrams}}. The S2:S1 fluorescence ratio allows quantitative comparison between the r-ETPA cross section and the classical one-photon absorption cross section. ICG is also important in medical imaging as the first FDA-approved near-IR contrast agent \cite{noauthor_is_nodate}. Although traditionally used as a one-photon dye, ICG will have further medical applications in deep tissue imaging if it exhibits significantly enhanced r-ETPA.

In this work, spectrotemporally resolved emission signatures of ICG are measured when photoexcited by broadband entangled photons in the near-IR region. No r-ETPA signals are measured within the detection limits of the measurements, allowing us to place an experimental upper bound on the r-ETPA cross section of ICG at $ 6 \times 10^{-23}$ cm$^2$/molecule. For comparison, the classical r-TPA cross section is also measured to be \(20(\pm13)\) GM where 1 GM equals $10^{-50}$ cm$^4$ s/molecule. The results suggest that having a resonant intermediate state does not significantly enhance the ETPA cross section of ICG in this case. The findings indicate that current measurement schemes with non-diffraction-limited focusing conditions, $<1\%$ fluorescence collection efficiencies, and $<10$ photons/s dark counts are not sufficiently sensitive for detecting r-ETPA signals in organic molecules with $<$ 20 GM classical TPA cross sections. The measured sequential absorption cross section for ICG is reported for the first time, and this may still be useful for future imaging and sensing applications.

\section{Methods}

\subsection{Measuring r-ETPA cross sections}

\begin{figure}[h]
\centering
\includegraphics[scale = 1.1]{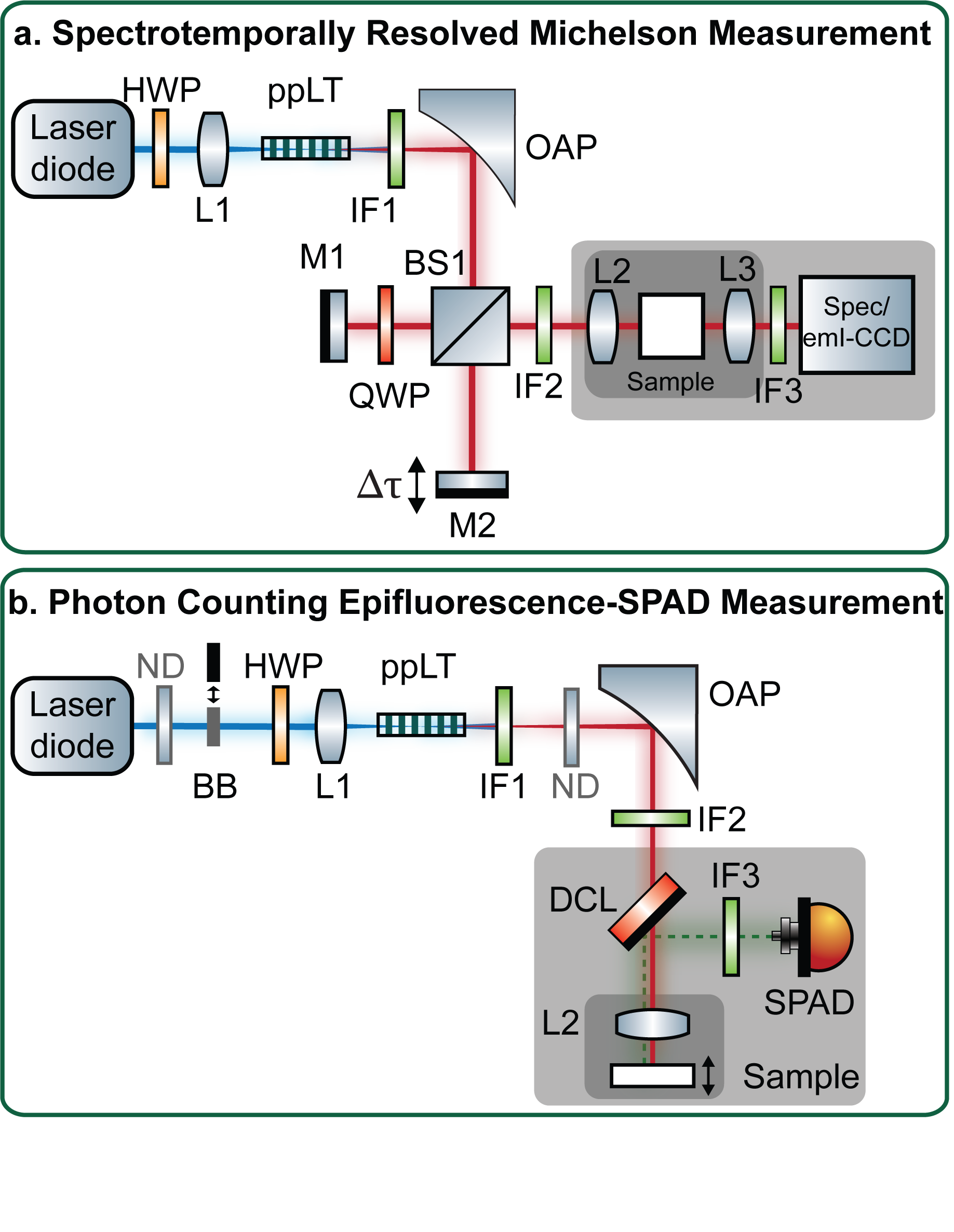}%
\caption{Experimental setups used in this work. a) A spectrotemporally resolved Michelson interference scheme with a grating spectrometer and emICCD. Fluorescence is collected along the direction of excitation beam propagation. b) A single-photon counting, epifluorescence scheme with a fiber-coupled SPAD as the detector. The gray shaded areas in both diagrams indicate physically sealed barriers that minimize scatter and ambient noise. HWP: half-wave plate, L: lens, ppLT: periodically-poled lithium tantalate, IF: interference filter set, OAP: off-axis parabolic mirror, BS: thin plate beamsplitter, QWP: quarter-wave plate, M: mirror, BD: beam dump, ND: neutral density filter wheel, BB: motorized beam block, DCL: dichroic longpass filter.
\label{fig:setup}}%
\end{figure}

A previously described continuous-wave (CW) laser-pumped, broadband entangled photon source is used in the experiments \cite{szoke_designing_2021, hickam_single-photon_2022}. Briefly, as shown in \textbf{Figure \ref{fig:setup}}, a 400-mW CW laser diode (Coherent) with a center wavelength of 403 nm is focused by a 400 mm plano-convex lens through a temperature-controlled Type 0 periodically poled lithium tantalate (ppLT) crystal. This produces collinear entangled photons centered around 806 nm with a bandwidth of $\sim$200 nm via spontaneous parametric down-conversion (SPDC). To minimize hot-band absorption \cite{mikhaylov_hot-band_2022} and one-photon scatter \cite{hickam_single-photon_2022}, edgepass filters are used to limit the bandwidth of the entangled photon excitation flux to 750-850 nm. Coincidence counting within the linear response range of the single photon avalanche diodes (SPAD, Laser Components) is used to verify that SPDC pairs rate scales linearly with pump power. The maximum SPDC pairs rate used here is $1.9 \times 10^{10}$ pairs/s, corresponding to 9.2 nW of pairs flux and an SPDC efficiency of $2.3 \times 10^{-10}$. 

Two different detection schemes are used to measure fluorescence from r-ETPA in ICG. Each scheme has its advantages and drawbacks. The first detection scheme, as shown in \textbf{Figure \ref{fig:setup}a}, uses a Michelson interferometer to introduce time delays ($\Delta \tau$) between the signal and idler photons in the excitation flux. A grating spectrometer and an electron-multiplying intensified charge-coupled device (emICCD, Princeton Instruments) measure any fluorescence signal. An achromatic lens with a 100 mm focal length focuses excitation flux down to $<$ 200 $\mu$m beam waist into a 10-mm thick cuvette. The cuvette holds an aqueous solution of 1 mM ICG. The second detection scheme (\textbf{Figure \ref{fig:setup}b}) is an epifluorescence scheme with a SPAD. Compared with literature \cite{tabakaev_spatial_2022}, a motorized beam block is also added to the pump path that blocks and unblocks the pump laser every 1 min to allow subtraction of dark counts and eliminate drift over long integration times. Also, instead of using different optical components to vary the pump and pairs rate, there are two locations along the beam path where a continuously variable metallic neutral density (ND) filter wheel is placed. The 2-mm thick ND wheel minimally affects the direction of beam propagation and introduces negligible dispersion. An aspheric lens with a 4.51 mm focal length and 0.55 numerical aperture (NA) focuses the excitation flux down to $<$ 35 $\mu$m beam waist into a sample cuvette, measured by a laser beam profiler (Femto Easy). The cuvette is 1 mm thick to reduce fluorescence reabsorption, and the solvent is changed to DMSO for improved quantum yield \cite{berezin_near_2007} while the dye concentration remains 1 mM.

The Michelson scheme eliminates false signals caused by uncorrelated photon pairs and one-photon events by analyzing the interference pattern in the spectrotemporally recorded signal \cite{hickam_single-photon_2022}. The disadvantage of the scheme is its loss and complexity. The nondeterministic splitting of the entangled pairs by the Michelson interferometer incurs a 75\% loss of visibility, and the spectrometer slit and grating induces another ~50\% loss. The emICCD's high dark counts ($1 - 1000$/s/pixel) add an extra source of noise and are not ideal for detecting single photons. Overall, the interferometer's long-term instability prevents acquisition times longer than a few minutes per data point. The main advantage of the epifluorescence and SPAD measurement is that it improves fluorescence collection efficiency by at least an order of magnitude to 1.5\% compared with the Michelson setup. Hours-long integration times per data point are also possible with the SPAD. The disadvantage is that the SPAD is a bucket detector and cannot discern false signals from one-photon scatter or residual pump leakage. Heavy optical filtering must instead be relied on. A total of 20 OD of optical filtering from stacked longpass filters is used to remove the 403 nm CW pump beam after SPDC. A total of 8 OD of shortpass filters are placed in front of the SPAD to minimize one-photon SPDC scatter into the detector. 

\subsection{Measuring classical r-TPA cross sections}

To benchmark the quantum enhancement of r-ETPA, the classical r-TPA cross section of ICG at 800 nm is first measured as it is not widely reported in the literature. Two-photon fluorescence from ICG is compared with that from a reference sample of R6G with known TPA cross sections \cite{makarov_two-photon_2008}. Both ICG and R6G are dissolved in DMSO and have the same concentration of 2.5 $\mu$M. The r-TPA cross section of ICG can be derived from the following formula \cite{deng_measurement_2020}:

\begin{eqnarray}
\eta_{ICG} \cdot \delta(\textrm{800 nm})_{ICG} = \eta_{R6G} \cdot \delta(\textrm{800 nm})_{R6G} \cdot \frac{F(\textrm{800 nm})_{ICG}}{F(\textrm{800 nm})_{R6G}} 
\cdot \left( \frac{P(\textrm{800 nm})_{R6G}}{P(\textrm{800 nm})_{ICG}} \right) ^2
\label{eq:one}
\end{eqnarray}

where \(\eta\) is the two-photon fluorescence quantum yield of ICG or R6G, \(\delta(\textrm{800 nm})\) is the TPA cross section of ICG or R6G at 800 nm excitation, \(F(\textrm{800 nm})\) is the measured two-photon fluorescence flux from ICG or R6G excited by an 800 nm laser, and \(P(\textrm{800 nm})\) is the excitation power used on ICG or R6G.

Classical two-photon fluorescence is measured by a calibrated, commercial Zeiss LSM 880 confocal laser scanning microscope with a 140-fs pulse width multiphoton tunable laser (Coherent). The sample solution is mounted into an imaging well formed by a 1-mm thick silicone spacer (CultureWell) securely placed on a glass slide and covered by a \#1.5 coverslip. The sample is then placed on the translation stage of the microscope, where a 40x C‐apochromat water immersion objective (1.2 NA) is used to excite the sample and collect fluorescence. The microscope is equipped with appropriate emission filters, a series of diffraction gratings, and 32 photomultiplier tubes for TPA fluorescence detection. The built-in photon counting mode of the microscope is used to measure the fluorescence count rate, and the measurement is averaged over 50 repetitions of 1 s integration time. The fluorescence flux \(F(\textrm{800 nm})\) values in \textbf{Equation \ref{eq:one}} for ICG and R6G are obtained by subtracting the background count rate of pure DMSO from the sample count rates. The excitation power \(P(\textrm{800 nm})\) ratio between R6G and ICG is calculated from the pulsed laser output powers, assuming linear losses of the laser power through the optical path before the sample plane. Specifically, to ensure the measured classical TPA fluorescence count rates are within the photomultiplier tubes' dynamic range, a \(\frac{P(\textrm{800 nm})_{R6G}}{P(\textrm{800 nm})_{ICG}}\) ratio of 0.5 is used since ICG fluorescence quantum yield is lower than that of R6G. The built-in spectral detection mode of the microscope (9 nm resolution) is also used to qualitatively measure the r-TPA fluorescence spectrum of ICG, with an optical filter cutoff at 647 nm.

\section{Results and Discussion}

\begin{figure}[h]
\includegraphics[scale = 0.45]{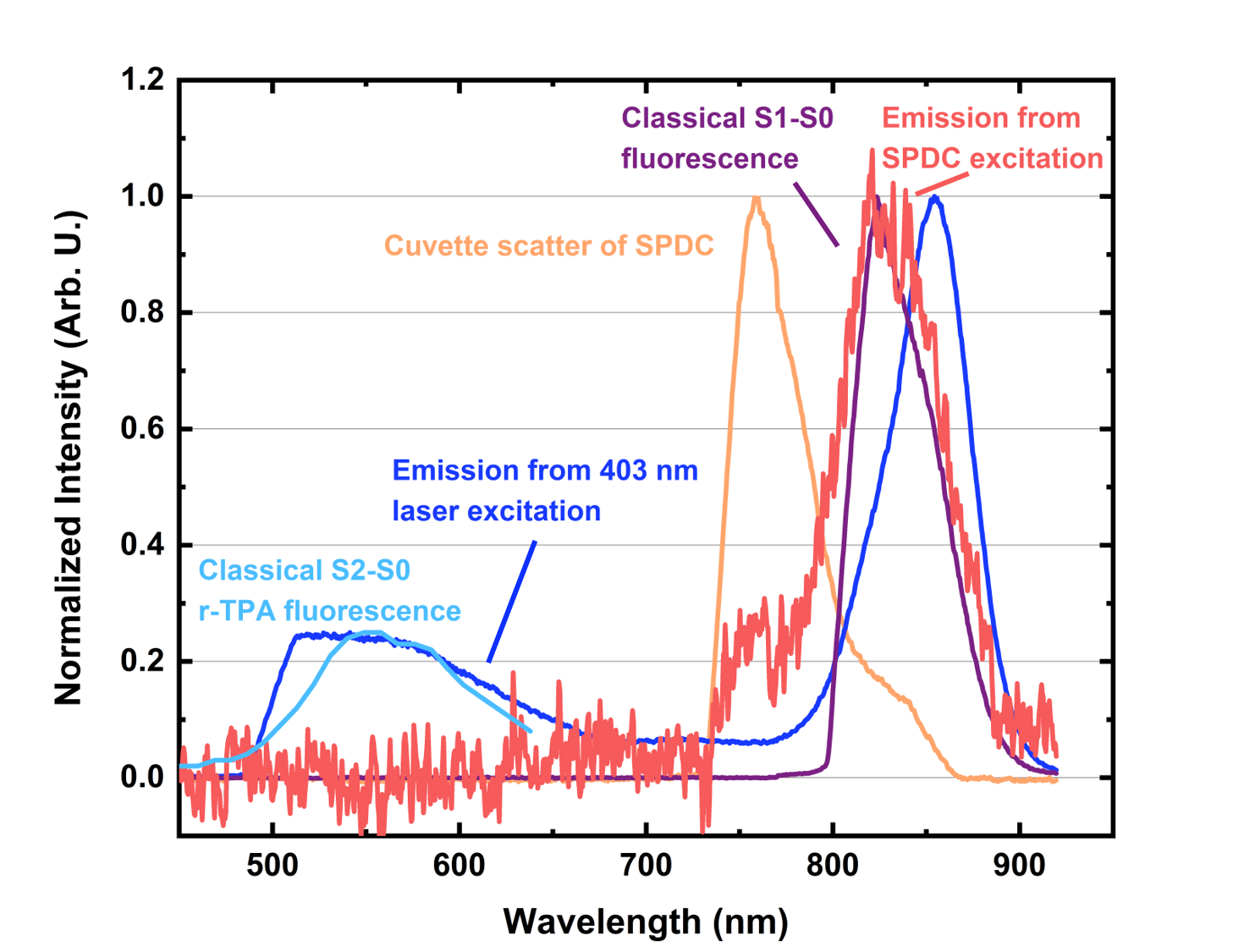}%
\caption{ICG fluorescence spectra when excited by an 800 nm pulsed two-photon laser (light blue), an 808 nm CW laser (purple), and the SPDC flux (red); also showing dual fluorescence from 403 nm CW laser excitation (deep blue), as well as cuvette front scatter of SPDC (orange). All spectra except the classical TPA fluorescence are taken using the epifluorescence measurement scheme (\textbf{Figure \ref{fig:setup}b}), with the fiber output sent to the spectrometer and emICCD. The classical TPA fluorescence spectrum is taken by the commercial two-photon absorption microscope and is cut off at 647 nm by an optical filter. These spectra are arbitrarily scaled and are not corrected for detector QEs.
\label{fig:icg-spectra}}%
\end{figure}

\textbf{{Figure \ref{fig:icg-spectra}}} compares the emission spectrum of ICG excited by the SPDC flux (red) with ICG's classical r-TPA S2-S0 fluorescence spectrum (light blue) and classical S1-S0 fluorescence spectrum (purple). The S2-S0 and S1-S0 emission peaks by 403 nm excitation (deep blue) are spectrally separated so their peak intensity ratio is used to estimate the S2-S0 fluorescence quantum yield of ICG for later calculation of the classical r-TPA cross section. Specifically, the intensity values at 555 nm (S2-S0 fluorescence peak) and 850 nm (S1-S0 fluorescence peak) are scaled by the quantum efficiencies (QEs) of the spectrometer grating \cite{noauthor_isoplane_nodate} and CCD camera \cite{noauthor_pi_nodate}. The ratio of the QE-corrected intensities is then multiplied by ICG S1-S0 fluorescence quantum yield of 0.12 in DMSO \cite{berezin_near_2007}, resulting in an estimated ICG S2-S0 fluorescence quantum yield of 0.02. The emission spectrum from SPDC excitation (red) at 20 s integration time does not have an apparent S2-S0 emission peak. Therefore, the more spectrally sensitive Michelson interference scheme (\textbf{Figure \ref{fig:setup}a}) is employed to examine if one- versus two-photon signals can be separated. 

\begin{figure}[h]
\centering
\includegraphics[scale = 0.53]{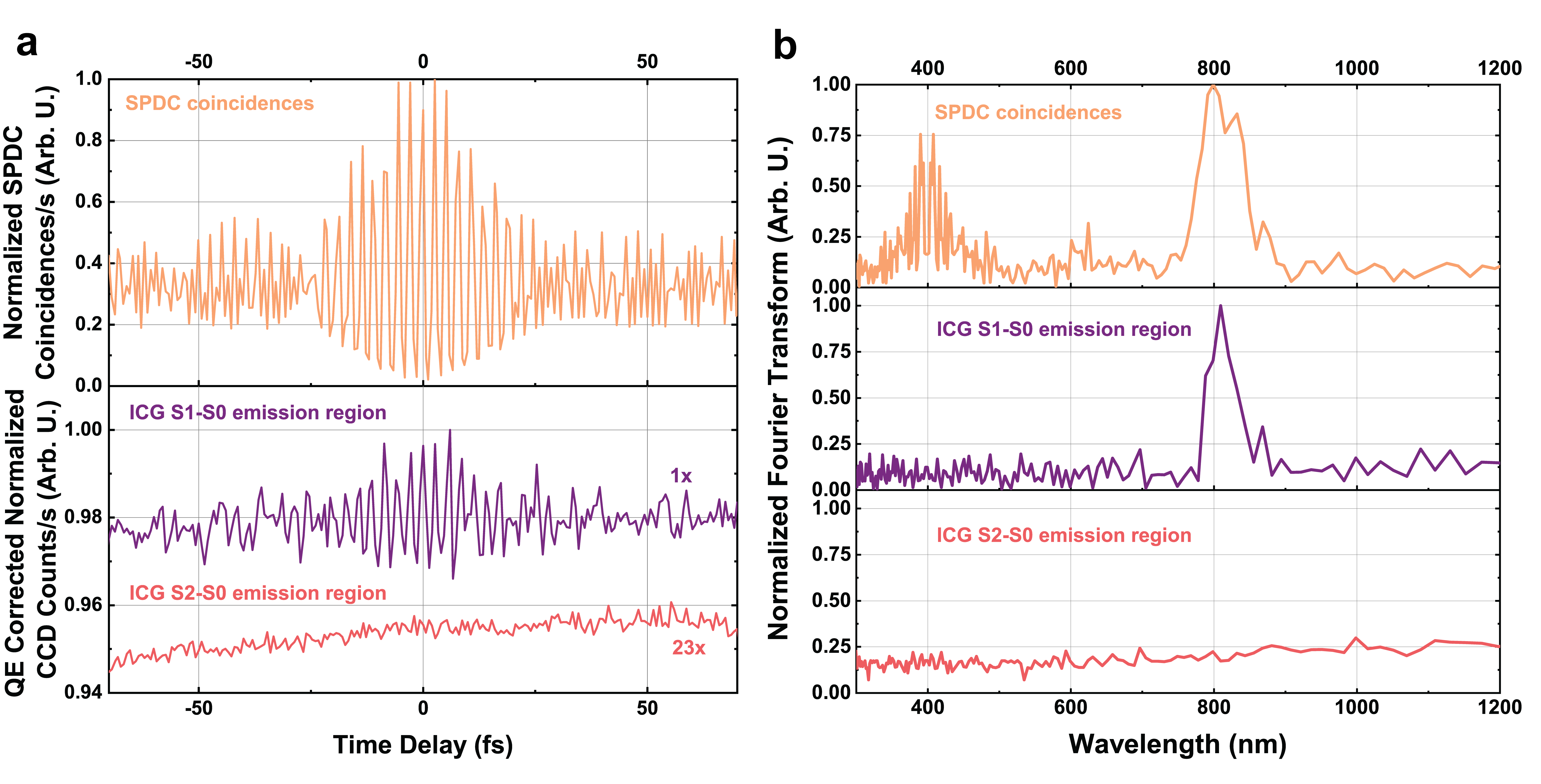}%
\caption{Spectrotemporally resolved, quantum efficiency corrected Michelson interferograms (a) and their Fourier transforms (b) of SPDC coincidences transmitted through a cuvette of water (orange), ICG's S1-S0 fluorescence region (purple), and ICG's S2-S0 fluorescence region (red). The Fourier transform of the interferometer pair rate exhibits a 2f peak at 403 nm, indicating frequency anti-correlation of the entangled photon state. In contrast, the Fourier transforms for ICG emissions do not have 2f peaks. No two-photon events are registered by the interferograms or the Fourier transforms, ruling out the presence of r-ETPA fluorescence from ICG.
\label{fig:michelson-data}}%
\end{figure}

\textbf{Figure \ref{fig:michelson-data}} compares the CCD interferograms of the ICG S1-S0 (purple) and S2-S0 (red) emission regions as a function of time delay. Coincidence rates from the SPDC flux, measured by a pair of SPADs and a timing circuit, are also shown in orange. Each data point at a certain time delay $\Delta\tau$ in the S1-S0 emission region (820-920 nm) represents a summed signal count accumulated over 10 s, and each data point in the S2-S0 emission region (450-650 nm) represents a summed signal count accumulated over 100 s. In the two-photon coincidence interferogram, for an entangled state, the peak amplitudes are larger than the valley amplitudes near time zero. In a TPA event, the interferogram of the fluorescence will follow that of the SPDC coincidences. For a one-photon absorption event, the interferogram's peak and valley components have equal amplitudes, which is the case for the ICG S1-S0 emission pattern \textbf{Figure \ref{fig:michelson-data}a}. On the other hand, the interferogram for the S2-S0 emission region is at or near noise level, indicating that no one- or two-photon signals exist within the spectral region despite this approach's higher signal-to-noise ratio compared with the spectral measurement of \textbf{Figure \ref{fig:icg-spectra}}. The conclusions are further reinforced by the Fourier transforms of the interferograms \textbf{Figure \ref{fig:michelson-data}b}, where only the SPDC coincidences have a strong 2f peak at $\sim$403 nm, and no peaks are measured in the S2-S0 emission region.

Finally, an epifluorescence and SPAD measurement scheme (\textbf{Figure \ref{fig:setup}b}) is used to confirm the lack of a two-photon induced signal at even longer integration times than possible with the Michelson scheme. The approach was previously used to successfully measure v-ETPA signals of R6G with 5+ hours of integration time \cite{tabakaev_spatial_2022}. Error analysis is first performed using 1 mM R6G to determine the optimal integration time that stabilizes the standard deviation of the SPAD measurements. Three hundred sets of repeated measurements are used. Within each set, SPAD counts are measured for 1 minute in the laser-off configuration, followed by 1 minute of measurement in the laser-on configuration. A mean counts/s value is then calculated from each 1-min bin. The 150 repeated laser-off and 150 laser-on measurements are then analyzed separately to confirm that each accumulated standard deviation of the mean is proportional to \(1/\sqrt{N}\) where \(N\) is the number of repeated 1-min measurements \cite{skoog_principles_nodate}. The data indicates that 50 min of integration time stabilizes time-dependent fluctuations and is still practical for performing multiple ETPA measurements.

\begin{figure}[h]
\centering
\includegraphics[scale = 0.5]{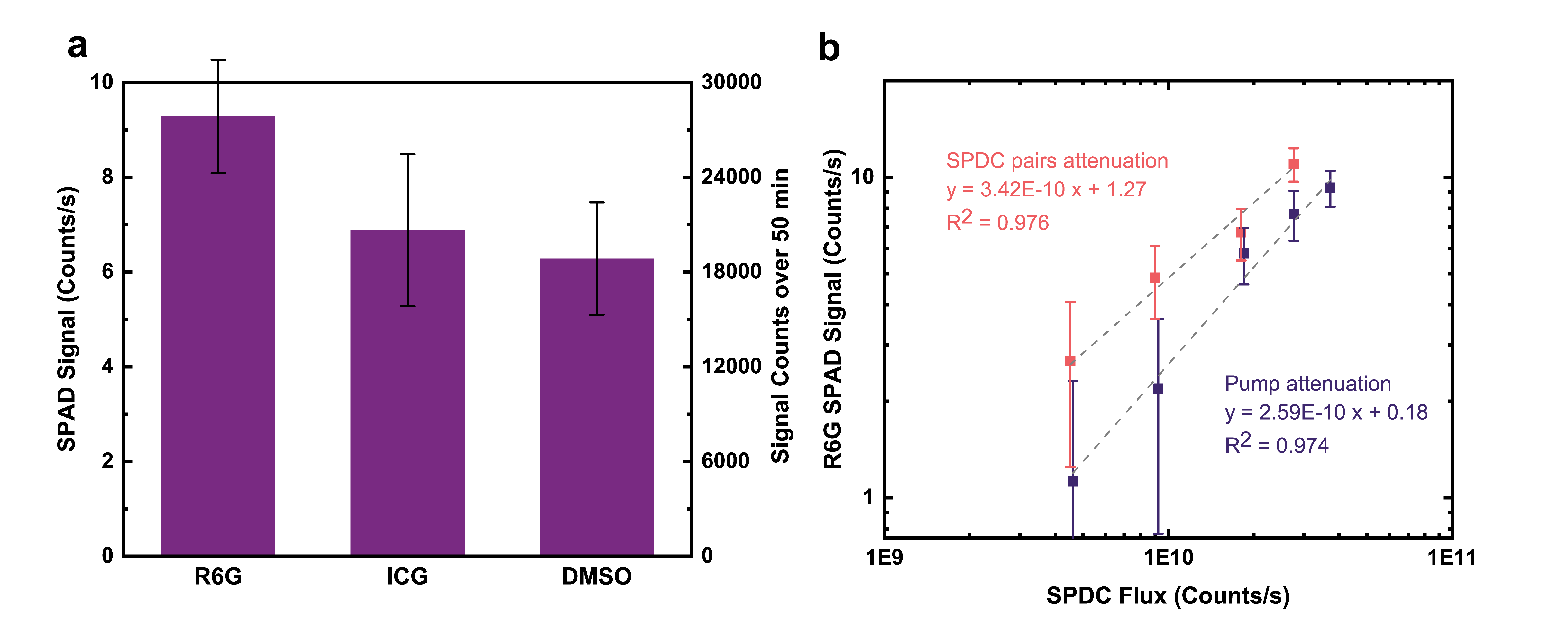}%
\caption{(a) SPAD signals of 1 mM R6G, 1 mM ICG, and DMSO, acquired over 50 min, in 50 repeated measurements of 1 min integration. Signal counts are calculated as laser-on counts minus laser-off counts (the total run time of each sample is, therefore, 100 min). The SPAD generates a readout every 100 ms. Each error bar is the standard deviation of the 50 measurements. ICG shows a background-level signal similar to that of DMSO, while R6G shows a statistically significant positive signal. (b) A pump versus pairs attenuation power study on R6G is conducted with the same integration times. Linear dependence on both pump and SPDC powers indicates one-photon processes such as hot-band absorption.
\label{fig:spad-data}}%
\end{figure}

\textbf{Figure \ref{fig:spad-data}a} shows a comparison between the SPAD counts from 1 mM R6G, 1 mM ICG, and solvent DMSO when excited by entangled photons. A wavelength range of 500-650 nm is selected by 8 OD of interference filters at the SPAD entrance, which encompasses both ICG’s S2-S0 r-TPA emission region and R6G’s S1-S0 v-TPA emission region.Although each signal in \textbf{Figure \ref{fig:spad-data}a} is already dark-count-subtracted, the solvent DMSO still shows positive counts, possibly indicating molecular Rayleigh scattering \cite{miles_laser_2001} of residual CW pump or SPDC. The ICG signal is statistically indistinguishable from the DMSO signal, confirming that r-ETPA is undetectable with an upper bound of $6 \times 10^{-23}$ cm$^2$/molecule for the cross section as later discussed. The R6G signal has a statistically significant signal above the solvent scatter. A pump versus pairs attenuation power study \cite{tabakaev_spatial_2022, mikhaylov_hot-band_2022, dayan_nonlinear_2005} using an ND filter wheel is performed to determine whether the signal is from one- or two-photon events. The measurement approach verifies ETPA when the fluorescence signal scales linearly with varying pump power but quadratically with pairs attenuation because the loss of one photon within an entangled pair destroys the entanglement.  However, \textbf{Figure \ref{fig:spad-data}b} shows that the R6G signal counts scale linearly with both pump attenuation and pairs attenuation, indicating the presence of one-photon processes such as hot-band absorption \cite{mikhaylov_hot-band_2022} instead of ETPA. While the pairs attenuation curve is shifted up with respect to the pump attenuation curve, we attribute this discrepancy as an artifact from the placement of the ND filter wheel in the beam path between pump and pair measurements. When moved, the ND wheel can alter background scatter at few photons per second levels. The results are, therefore, too close to the instrument noise floor to draw statistically sound conclusions about ETPA fluorescence in either ICG or R6G.

Experimental upper bounds can be placed on the r-ETPA cross section (\(\sigma_{ETPA}\)) of ICG based on the two measurement schemes, as summarized in \textbf{Table \ref{tab:table1}}. Each scheme’s excitation flux, focusing conditions, collection efficiency, and the optical properties of ICG are used in the calculation \cite{hickam_single-photon_2022}. Additionally, the classical r-TPA cross section of ICG (\(\delta_{TPA}\)) is calculated to be \(20(\pm13)\) GM according to \textbf{Equation \ref{eq:one}}, and is included in the table for comparison. The measured \(F(\textrm{800 nm})_{ICG}\) value is \(1899 (\pm836)\) Hz, and the \(F(\textrm{800 nm})_{R6G}\) value is \(22365 (\pm429)\) Hz. To arrive at the \(\delta(\textrm{800 nm})_{ICG}\) value, the following approximations are made: 1) because reported TPA cross sections of R6G vary by up to 1 order of magnitude, regardless of solvent and excitation conditions \cite{makarov_two-photon_2008, oulianov_observations_2001}, \(20 (\pm10)\) GM is used as an estimate for \(\delta(\textrm{800 nm})_{R6G}\); 2) \(\eta_{ICG}\) is estimated as 0.02 as explained earlier; 3) \(\eta_{R6G}\) is approximated to be 0.95 \cite{kubin_fluorescence_1982, magde_fluorescence_2002}. Because of the approximations, the calculated r-TPA cross section of ICG is emphasized as an order-of-magnitude estimation. 

\begin{table*}
\caption{\label{tab:table1}Summary of classical TPA cross sections and detection limits for ETPA cross sections.}
\begin{ruledtabular}
\begin{tabular}{cccccc}
 &&\multicolumn{2}{c}{Michelson scheme detection limits}&\multicolumn{2}{c}{SPAD scheme detection limits}\\
 Molecule&Classical \(\delta_{TPA}\)&\(\sigma_{ETPA}\)&Equivalent \(\delta_{TPA}\)\footnote{Calculated from excitation fluxes and focusing areas.}&\(\sigma_{ETPA}\)&Equivalent \(\delta_{TPA}\)$^{\text{a}}$\\ 
 &(GM)&(cm\(^2\))&(GM)&(cm\(^2\))&(GM)\\ 
 \hline
 ICG&\(20(\pm13)\)\footnote{Measured relative to R6G \(\delta_{TPA}\) with 800 nm excitation.}
 &$9 (\pm2)\times 10^{-23}$&$2 (\pm2)\times 10^{13}$&$6 (\pm2) \times 10^{-23}$&$3 (\pm2)\times 10^{12}$\\
R6G&\(20(\pm10)\)\footnote{Based on literature \cite{makarov_two-photon_2008, oulianov_observations_2001}.}
 &$4 (\pm2)\times 10^{-24}$&$8 (\pm2)\times 10^{11}$&$8 (\pm2) \times 10^{-24}$&$4 (\pm2)\times 10^{11}$
\end{tabular}
\end{ruledtabular}
\end{table*}

Using the excitation fluxes and focusing areas of the two measurement schemes, the r-ETPA cross section detection limits can be converted \cite{noauthor_two_nodate} to equivalent \(\delta_{TPA}\) values of $2 (\pm2)\times 10^{13}$ GM and $3 (\pm2)\times 10^{12}$ GM, as shown in \textbf{Table \ref{tab:table1}}. Comparing these values with ICG's classical r-TPA cross section of \(20(\pm13)\) GM, we conclude that having a resonant intermediate state must enhance ICG's ETPA cross section by less than 11 orders of magnitude. The conclusion is supported by previous theories \cite{oka_two-photon_2018, li_enhancement_2023} suggesting that ETPA would only offer up to 4 orders of magnitude enhancement on the excited state population in atomic systems, and r-TPA may only be $\sim$1.7 times more prone to quantum enhancement than v-TPA when excited by 100 nm bandwidth of SPDC photons as used in our work \cite{oka_two-photon_2018}. The determined cross section upper limits for R6G are also listed in \textbf{Table \ref{tab:table1}}. These detection limits agree with theoretical predictions \cite{landes_quantifying_2021} and previous experiments \cite{parzuchowski_setting_2021, hickam_single-photon_2022}, suggesting that the measured upper bounds on ICG in this work are reasonable. The ICG detection limits differ from the R6G detection limits by $\sim$1 order of magnitude mainly because ICG's absorption bands are broad and heavily overlap with emission bands, leading to a higher chance of fluorescence reabsorption. ICG also has a lower quantum yield than R6G. Note that the correlation time of the entangled photons is $<$ 50 fs (\textbf{Figure \ref{fig:michelson-data}a}), which is much shorter than the sub-ns lifetime of ICG \cite{berezin_near_2007}. Therefore, if present, ICG's S1-S2 transition is not significantly hindered by the decay of the intermediate state. 

The results are intended as upper bounds specific to the described experimental configurations, which have proven helpful in other studies of ETPA \cite{hickam_single-photon_2022, parzuchowski_setting_2021}. In the future, if near UV to deep UV entangled photons can be created, molecules with higher S2-S0 quantum yields can be studied, such as azulenes, aromatic acenes, polyenes, and metalloporphyrins \cite{itoh_fluorescence_2012}. Earlier in the field, theories have predicted that maximizing the bandwidth of the entangled photon source, which in turn maximizes the brightness and minimizes the entangled photon correlation time, would be the best approach to increasing ETPA efficiency \cite{oka_two-photon_2018, dayan_nonlinear_2005, landes_quantifying_2021}. However, experimentally, the broad momentum matching cone is difficult to collimate, leading to focused spot sizes that are still 1-2 orders of magnitude away from the diffraction limit \cite{szoke_designing_2021}. Furthermore, maintaining the single-photon-per-mode limit in nonlinear light-matter interactions may not even be critical. Bright squeezed vacuum, a class of SPDC fluxes that saturate the single-photon-per-mode limit, also exhibits many quantum-enhanced properties in two-photon absorption \cite{georgiades_nonclassical_1995, dayan_two_2004}, optical harmonics generation \cite{spasibko_multiphoton_2017}, and ultrafast spectroscopy \cite{cutipa_bright_2022}. Future experiments using bright squeezed vacuum sources generated in a single, waveguided spatial mode could further improve the detection limits for entangled multiphoton processes.

\section{Conclusion}

We investigated the validity of r-ETPA in ICG and placed upper bounds on the possible enhancement. We measured the spectrotemporally resolved emission signatures of ICG excited by broadband entangled photons in near-IR. Similar to many reported v-ETPA measurements, no r-ETPA signals are measured, with an upper bound for the cross section placed at $6 \times 10^{-23}$ cm$^2$/molecule. In addition, the classical r-TPA cross section of ICG at 800 nm is measured to be \(20(\pm13)\) GM. The same measurements are performed on R6G for bench-marking. The results suggest that having a resonant intermediate state does not significantly enhance the ETPA cross section of ICG beyond the detection limits of this study. The findings indicate that current measurement schemes with non-diffraction-limited focusing conditions, $<1\%$ fluorescence collection efficiencies, and $<10$ photons/s dark counts are not sufficiently sensitive for detecting r-ETPA signals in organic molecules with small classical TPA cross sections. Further improvements to lower the instrument detection limits will be the key to the successful detection of r-ETPA as well as v-ETPA signatures.

\begin{acknowledgments}
This material is based upon work supported by the U.S. Department of Energy, Office of Science, Office of Basic Energy Sciences, under Grant DE-SC0020151 (S.K.C.). 
\end{acknowledgments}

\section*{Conflict of Interest Statement}

The authors have no conflicts to disclose.

\section*{Author Contributions}
\textbf{Manni He}: Conceptualization (equal), Data Curation (lead), Formal Analysis (lead), Methodology (equal), Resources (lead), Software (equal), Writing/Original Draft Preparation (equal), Writing/Review \& Editing (equal).

\textbf{Bryce P. Hickam}: Conceptualization (supporting), Data Curation (supporting), Formal Analysis (supporting), Methodology (equal), Resources (supporting), Software (equal).

\textbf{Nathan Harper}: Formal Analysis (supporting), Methodology (supporting), Resources (supporting), Writing/Review \& Editing (equal).

\textbf{Scott K. Cushing}: Conceptualization (equal), Funding Acquisition (lead), Methodology (supporting), Writing/Original Draft Preparation (equal), Writing/Review \& Editing (equal).

\section*{Data Availability Statement}

The data that support the findings of this study are available within the article. Additional data are available from the corresponding author upon reasonable request. 

\bibliography{paperreferences}

\end{document}